# Meta-Analysis of the Accuracy of Syndromic Surveillance

## Liaquat Hossain and Derek Kham

Nebraska Healthcare Collaborative Chair of Population Health, Department Chairperson, Department of Cyber Systems, College of Business and Technology, The University of Nebraska at Kearney, NE 68849, Email: hossainl2@unk.edu

Senior Research Assoc, Division of Information & Tech Studies, University of Hong Kong, HK

**Abstract**
We present the first meta-analysis of co-evolutionary learning networks for digital disease surveillance research over last 10 years. In doing so, we show the co-evolution and dynamical changes that occurred in academic research related to digital disease surveillance for improving accuracy, approach and results. Using dynamic network analysis, we are able to show the incorporation of social media-based analytics and algorithms which have been proposed and later improved by other researchers as co-evolutionary learning networks. This essentially demonstrates how we improve our research and increase accuracy through feedback loop for correcting the behaviour of an open system and perhaps infer learning patterns, reliability and validity using 10 years scientific research in digital disease surveillance.

**Introduction**
Digital disease surveillance can be referred as sharing and capturing informal electronic information which have impact in reducing the time to recognize an outbreak and therefore, offers opportunity for governments and communities to be proactive in mobilizing resources for public health response (Brownstein, et al., 2009). Studies have drawn on a diversity of sources including search queries data (*e.g.,* Google and Yahoo), social network data (*e.g.,* Tweeter microblogs) and online health information (*e.g.,* Wikipedia articles) access data. However, there has been a significant debate about accuracy, robustness and precision of investigative approaches and results for developing public health surveillance, which harnesses the web for the past 10 years. A significant number of research appeared in leading journals and professional fields in proposing and subsequent validation of accuracy, robustness and precision of public health surveillance by harnessing the web. The first study of web access logs and influenza was published in 2004 (Johnson et al, 2004). Google Flu Trends was introduced in 2008 (Ginsberg et al., 2009), with major updates in 2009 (Cook et al., 2011) and 2011 (Copeland et al., 2013). Google Flu Trends was criticized for missing the 2009 pandemics and for 2013 overshot (Lazer et al., 2014). Most digital disease surveillance studies obtain high to very high correlations with traditional clinical surveillance data (Figure 4; e.g., in Ginsberg et al 2009, GFT at 0.94 with US sentinel surveillance influenza-illness-like (ILI) data). Nonetheless, correlations might not capture absolute errors and particular deviations, if any. Recommendations of studies vary widely. We explore the effects of assimilating GFT into more complex surveillance models for accuracy, approach and results of 10 years digital disease surveillance research.
,
We present a meta-analysis of 10 years' research in digital disease surveillance aiming to: 1) provide synthesis for complete coverage of all relevant studies; 2) look for the presence of heterogeneity; and, 3) explore the robustness of the main findings using dynamic network analysis. This is achieved through a systematic review of all relevant research for presenting a balanced and impartial summary of the research. The overall goal is to develop capability, through collection and analysis of big open data, for detecting the spread and outbreak of infections using Internet search queries and social media data. The work was carried out in three phases: 1) review of the evolution and co-evolution of approaches to outbreak detection algorithms; 2) systematic review of study designs; and, 3) meta-analysis of study results for showing how co-evolutionary learning networks evolve in improving accuracy, approach and results. The detailed design of our study is based on the development of dynamical co-evolutionary networks using the meta-analysis framework proposed in PRISMA (Preferred Reporting Items for Systematic Reviews and Meta-Analyses, which is a widely-acknowledged as evidence-based minimum set of items for reporting in systematic reviews and meta-analyses (PRISMA, 2009). Our study presents the meta-analysis of 10 years of research consisting of 65 influential research articles in digital disease surveillance, categorized by: internet and clinical data sources, keywords selection and language processing, mathematical assimilated modelling, limitations raised and improved, as well as the studies' objectives and recommendations.

**Co-evolutionary Networks of Digital Disease Surveillance**
Figure 1a and 1b below presents a system dynamic model (SDM) of the co-evolution of computational algorithms for internet influenza surveillance, incorporating 1) PRISMA 2009 checklist (PRISMA, 2009), and ii) double-loop learning model. The basic SDM framework consists of four components: input, process, output and feedback. PRISMA items are matched to each component and placed in *blue* boxes. Green boxes represent the 4 system dynamic components: inputs, processes, outputs and feedbacks. Blue boxes are standard PRISMA items. For our purposes, two items were added (in lighter blue): "modeling" is specific to digital disease surveillance algorithms; "recommendation" reflects the overall conclusion of each study. The SDM incorporates double loop learning, where feedback loop 1 corrects the system while keeping in with the given goals, values, plans and rules; feedback loop 2 corrects the system by modifying the governing variables (see Argyris and Schön, 1978). In addition to the above basic structure, the possible values of each parameter are also listed (white boxes). For brevity, not all parameters on limitations are shown. The model provides the basis for our assessment criteria of the studies, and for quantification of parameters for the statistical and network analyses.

Figure 1a. SDM of the co-evolution of digital disease surveillance

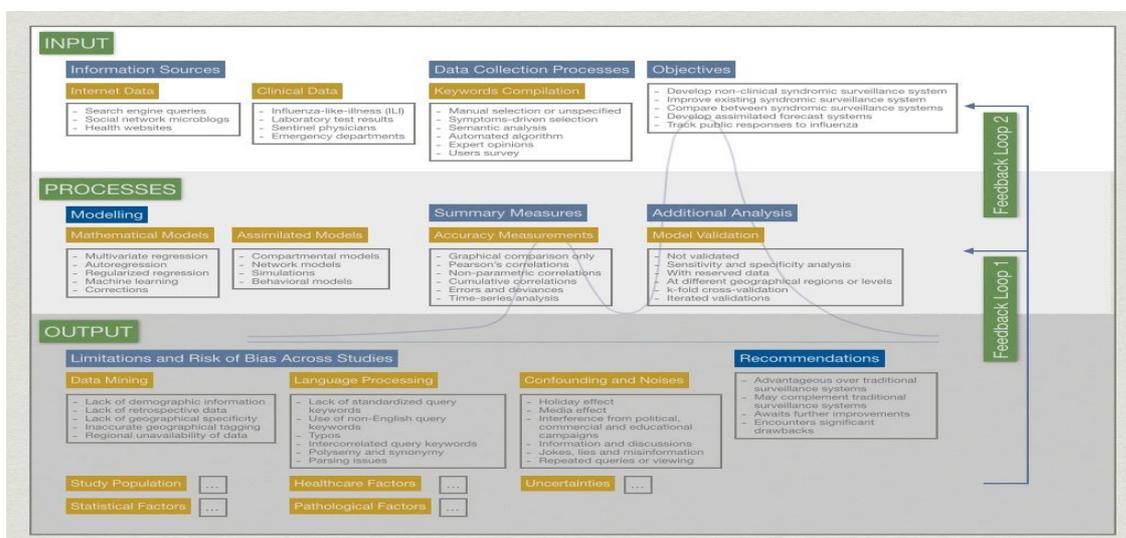

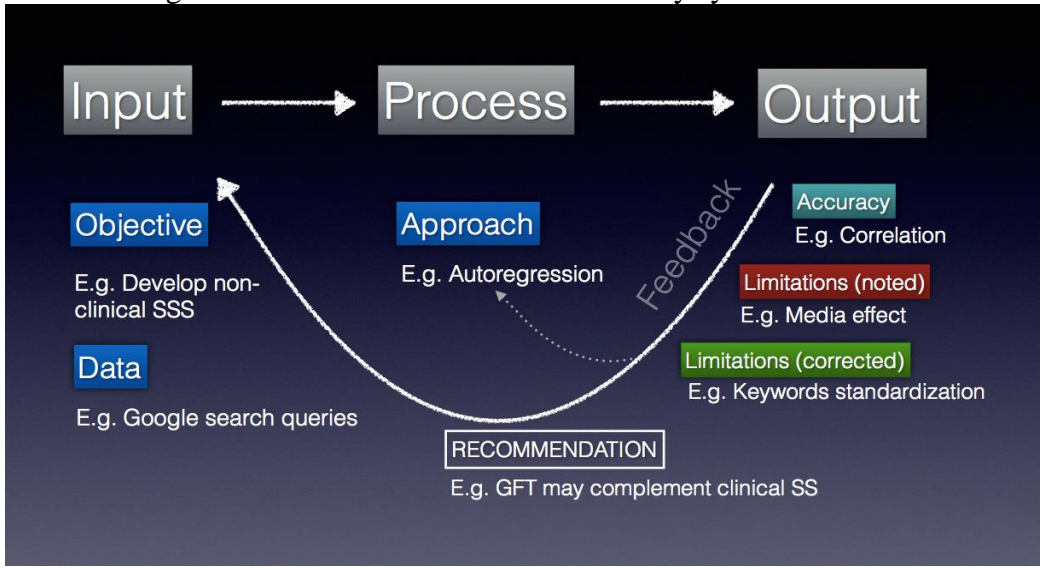

Figure 1b. 1a. SDM of the co-evolutionary systems model

Three steps were involved in developing the model. First, we selected PRISMA items (*blue* boxes) that are relevant to the development of computational algorithms and assigned them under the SDM components. Second, we constructed parameters for each included PRISMA item with respect to the computational algorithms (*yellow* boxes). Third, after an initial reading of the articles by the authors, a set of possible values (*bullet* points) were identified for each of the parameters. For example, for the parameter "accuracy measurement", six values were identified: "1 graphical comparison only", "2 parametric correlation", "3 non-parametric correlations", "4 errors and deviances", "5 time-series analysis" and "6 cumulative correlations".

**Data Sources and Categorization using PRISMA**

Journal articles on internet influenza surveillance were obtained from the following 14 general and biomedical databases: PubMed, Web of Science, Google Scholar, Cochrane, WHO GHL, Embase, Global Health, SciELO, CINAHL, BNI, IndMed, Trip, AMED, and Health Business Elite. The following combinations of keywords were used: 1) "influenza" + "internet", 2) "influenza" + "google flu trends", 3) "influenza" + "google trends", 4) influenza" + "Twitter".

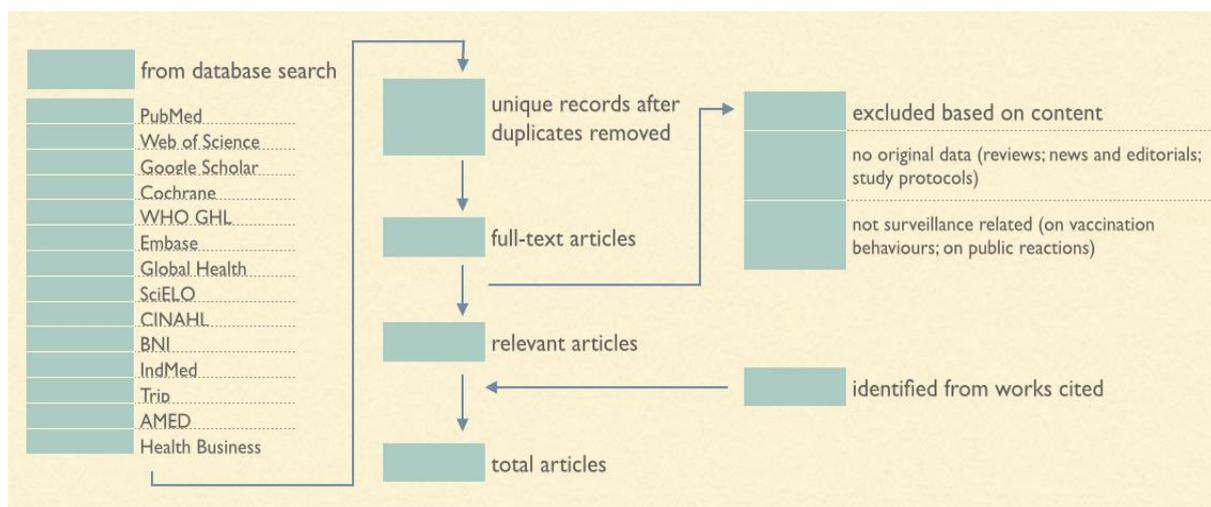

Figure 2. Flowchart of article search and selection process

Figure 2 above highlights the conceptualization of 6 iterative process used in search and selection processes of our study: 1) a database search was followed by duplicates removal; 2) full-text articles were screened; 3) exclusion of articles without original data, such as reviews, news and editorials and study protocols; 4) articles about influenza and the internet, which do not include studying disease surveillance, such as study of vaccination behaviors and public reactions only were also excluded; 5) further articles were identified by checking citations in relevant articles; and, 6) where a conference paper is highly similar in content to another journal article or conference paper by the same author(s), only the journal article or the earlier conference paper was included.

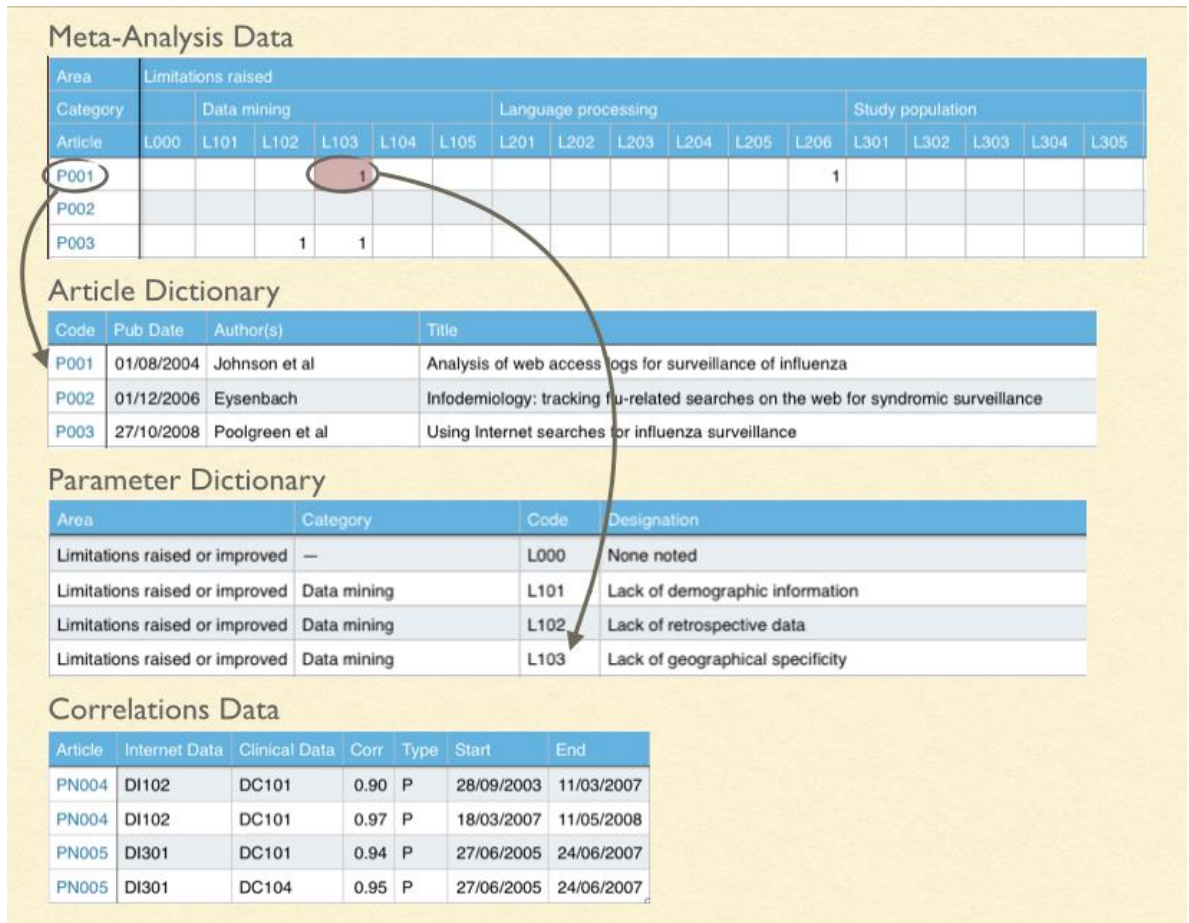

Figure 3. Illustration of data structure and coding system used in analysis

A database structure and coding system presented in Figure 3 were designed according to the above model, into which data extracted from the papers were entered. The highlighted datum indicates that the article Johnson et al 2004 raises the limitation that data collection for digital disease surveillance lacks geographical specificity. Also shown is an example data structure of the correlations data (bottom table). Each row records a correlation reported in an article between results from internet and clinical surveillance, alongside with the article code (e.g., PN004, which is the Ginsberg 2008 paper) and data codes (e.g., DI102 for sentinel physicians ILI data), start and end dates of data included, and the type of correlation measured (e.g., Pearson's correlation ("P")). For simplicity, only shows a few lines and columns of each table are shown.

Data manipulations, statistical analysis and plotting of bar graphs were carried out in Stata 13.1. Social network analyses and network diagrams were performed and produced using UCINET 6.554. Initially, bar graphs were made for studying the raw distributions of parameters over time. Then, the relationships among articles for the parameters were quantified using standard centrality measurements (e.g. eigenvectors, betweenness and closeness). The basic idea is that when two articles share the same parameter (e.g., identifying the same study limitation), their relatedness increases by 1. The results were encoded into adjacency matrices for producing network diagrams. Furthermore, reported correlations between digital disease surveillances and traditional clinical surveillances were extracted from the articles, including details of locations, data sources, time period(s), correlation type (e.g., parametric or non-parametric) and values. In figure 4 below we present the rough plot of the correlation outcomes reported from 35 internet influenza surveillance studies, with data included from 2004-2014. Left: studies using Google Flu Trends or Google Trends search query data. Right: studies using Twitter microblogs or Wikipedia search queries data. Each outcome compares between an internet influenza surveillance method (e.g., Google Flu Trends) and a clinical standard (e.g., sentinel physician's data on influenza-like-illnesses (ILI)). ILI: influenza-like illness; SP: sentinel physician's data; Lab: laboratory confirmed cases. Solid lines: Pearson's correlations (or unspecified); dotted lines: Spearman's rank correlations. Notice that the vertical scales of the graphs have been adjusted for clearer presentation.

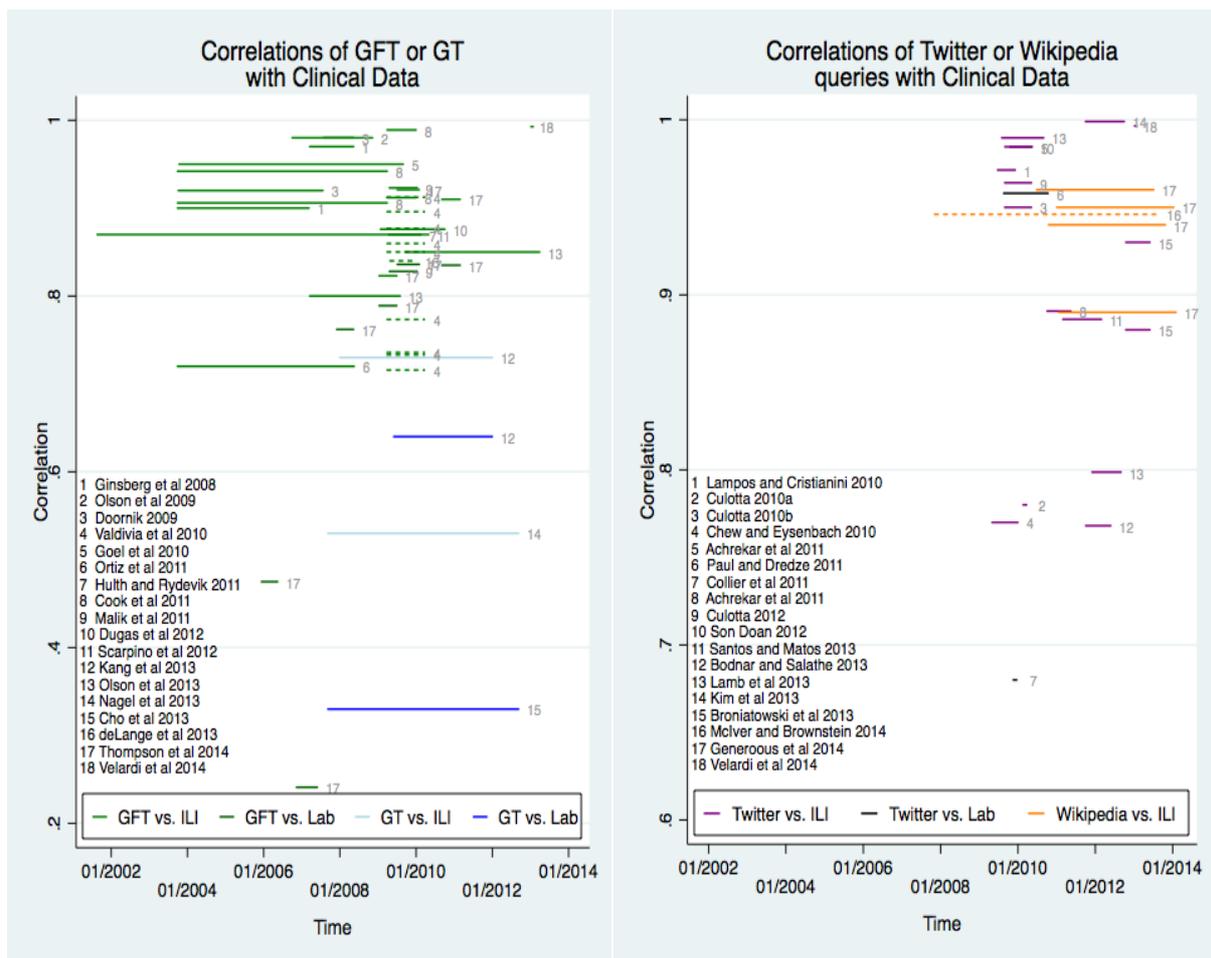

Figure 4. Rough plot of correlation reported in influenza surveillance studies from 2004-2014

**Results of Co-evolutionary Learning Networks**

In order to assess feedback loop 1 (which corrects the system while keeping in with the given goals, values, plans and rules), we examined the limitations raised and improved by the studies over time, in broad categories (see Figures 5-7). There was a gradual increase in the total number of limitations being raised (Figure 5), mostly on confounding factors and noises (*e.g.,* third-person information and discussions on the Internet). However, in 2013 and 2014, there was a sharp increase in limitations raised in the uncertainties category, consisting in large part of recognition about the lack of transparency of Google's GFT algorithm. Improvements on limitations appear to occur with a delay after limitations were raised: while the first digital disease surveillance studies appeared since 2004, improvements were made on the limitations starting in around 2009. Most are on confounding and noises, with a focus on correcting for media and calendar effects. In figure 5 below, we present the bar graphs of limitations raised (left) and improved (right) by the articles vs. time. Uncertainties due to the lack of transparency of the GFT algorithm were often noted in articles from recent years (within the "uncertainties" category, $p_{trend}=.0006$). There is also an increasing demand for accurate depictions of internet user populations' compositions and behaviors (under the "study population" category, $p_{trend}=.042$). Most of the limitations improved are about confounding and noises (light green bars), which involve mostly methods for correcting media and calendar/holiday effects.

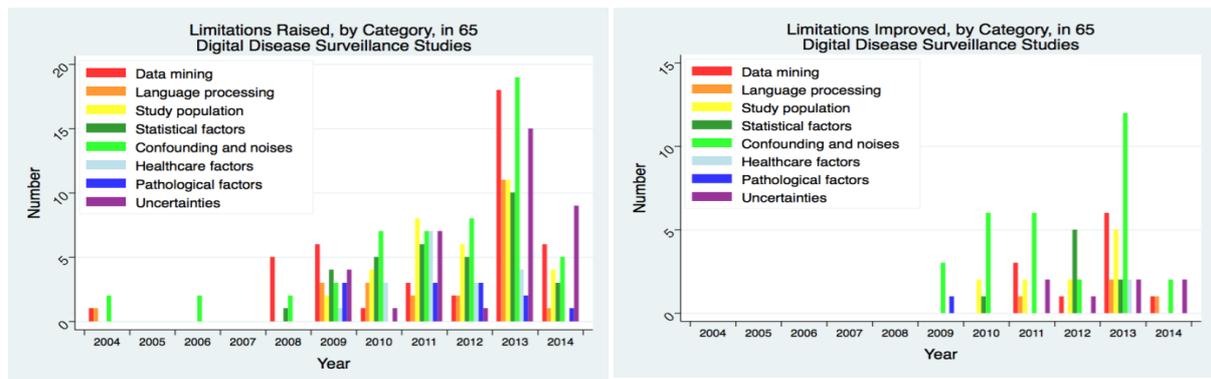

Figure 5. Bar graphs of limitations raised (left) and improved (right) by the articles vs. time

We further analyzed and visualized the relationships among the papers at four times frame: from 2004 up until 1) 2011, 2 2012, 3) 2013 and 4) 2014 (see figures 6 and 7 below). The basic idea is that two papers are related at 1 degree if they raised the same limitation, and at 2 degrees if they raised two of the same limitations, and so on; similarly for papers improving on the same limitation(s). For limitations raised, sharp increases in ties are observed in years 2013 and 2014, suggesting sharp increases in influences and/or collaborations amongst the authors. These include, from roughly 2012 to 2013, approximately doubling of the overall degree centralization of and number of clique structures in the network, halving in closeness, as well as gradual increase in betweenness and decrease in eigenvector of individual papers in the network. For limitations improved, similar trends were observed, including doubling in degree centralization from roughly 2012 to 2013. However, while the number of cliques increased less sharply, the increase in betweenness was much more drastic, suggesting that influences and/or collaborations were merely starting to happen then.

Figures 6 and 7 also illustrate the distribution of limitations raised and improved over the three main existing kinds of digital disease surveillance methods: Google search queries, Twitter microblogs, and health websites (e.g. Wikipedia) access logs. In terms of limitations raised, Figure 7 shows that the several kinds of studies were initially quite mixed, but became more segregated over time. This suggests that even though a certain level of collaboration and influence among the groups was apparently maintained (such as on raising more generalized limitations), over time the studies were raising limitations that are more specific and exclusive to the kind of surveillance method in question. In contrast, Figure 8 shows that limitations improved started clearly as two main groups: one for Google studies and another for Twitter studies. The two groups were joined together over time, with Twitter studies dominating the core of the network. One explanation could be that due to the availability of richer expressions and social networks information in microblog discussions, compared with the succinctness of search engine queries entered mostly by individual users, microblog studies might allow more flexible room for limitations improvement than search engine query studies.

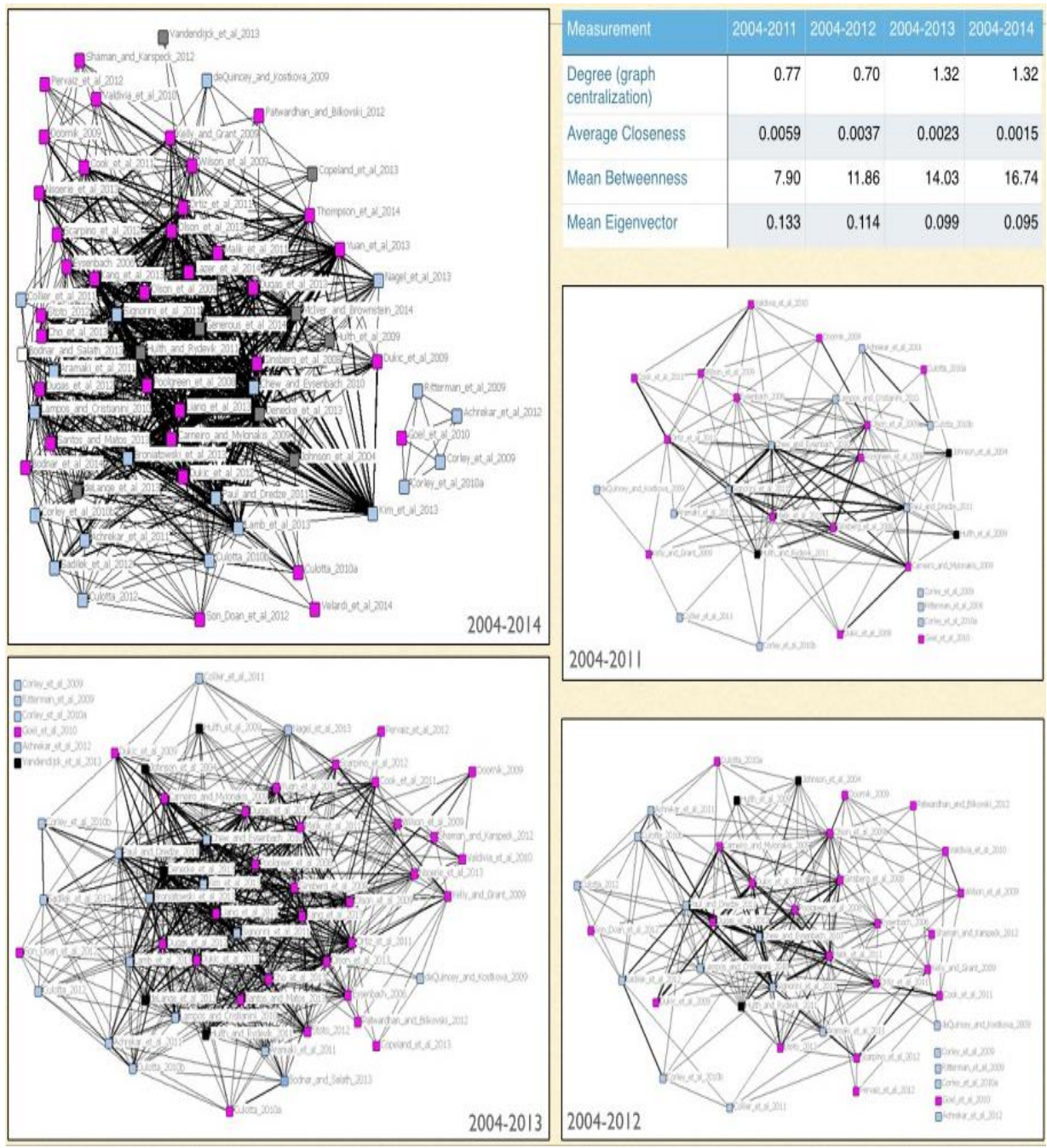

Figure 6. Network depiction of articles raising the same limitation(s) from 2004-2014

The issue of transparency in the GFT algorithm also remains to be solved, which might be restricting improvements. Figure 6 below provides network depiction of articles raising the same limitation(s) over time: a) 2004-2011, b) 2004-2012, c) 2004-2013, d) 2004-2014. Each node represents one article. The thickness of the tie between each two node corresponds to the number of identical limitations raised by them. Papers with high centrality can be seen, e.g. Polgreen et al (2008), Ginsberg et al (2009) and Yuan et al (2013). The unconnected nodes are papers raising limitations that are not subsequently also raised by other papers. Summary centrality measurements are presented alongside in the table. Node colours: pink - Google Flu Trends/Google Trends studies; blue - Twitter studies; black - Wikipedia or other health website studies. Figure 7 below provides network depiction of articles improving on the same limitation(s) over time: a) 2004-2011, b) 2004-2012, c) 2004-2013, d) 2004-2014. Node colours: pink - Google Flu Trends studies; blue - Twitter studies; black - Wikipedia or other website studies. (Isolates are not shown.) The period transiting from 2012 to 2013 shows a marked increase in the number of same limitations being raised, as seen in the increase in degree and betweenness and the decrease in closeness among the nodes.

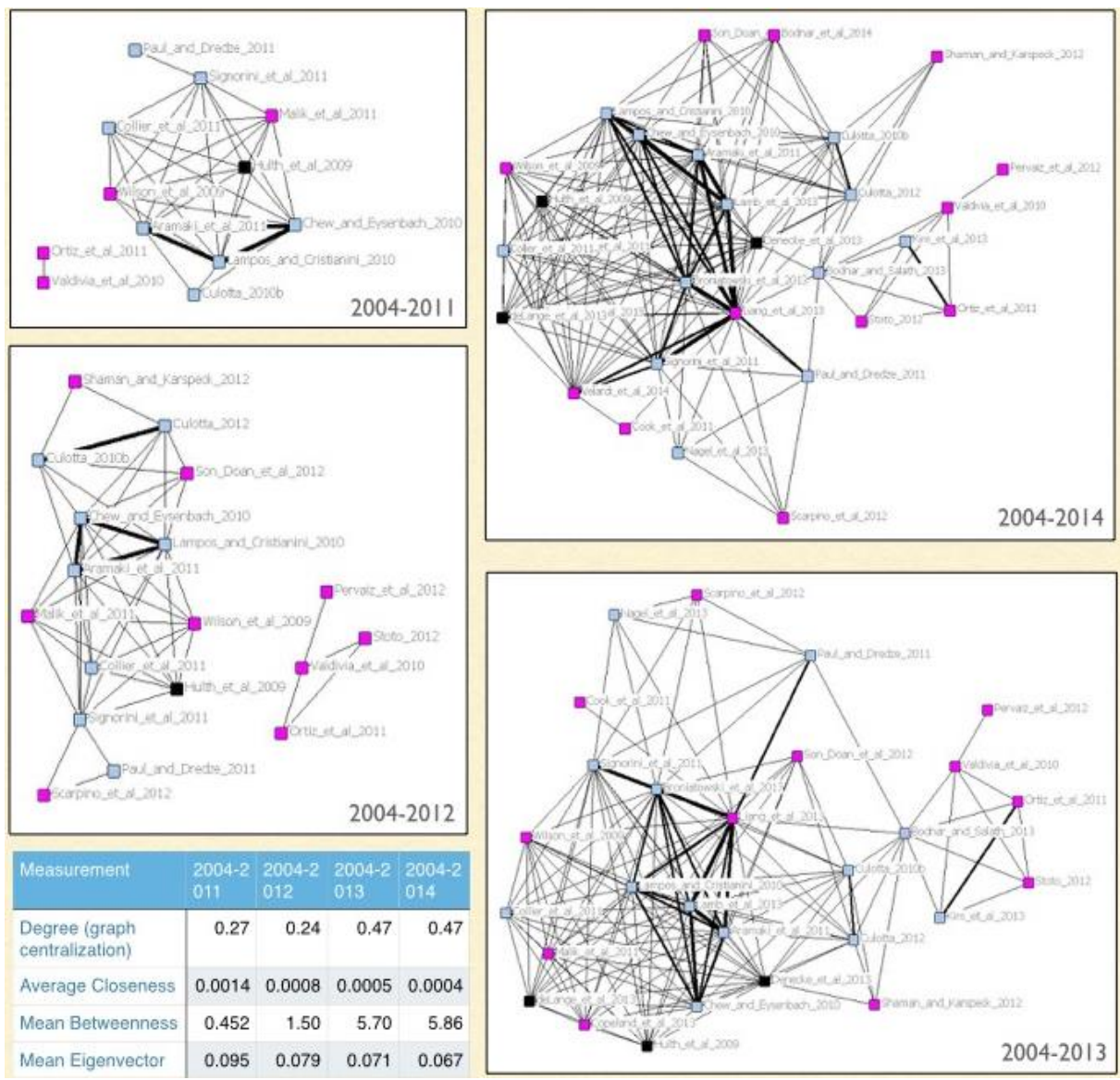

Figure 7. Network depiction of articles improving on the same limitation(s) from 2004-2014

In order to assess feedback loop 2, which corrects the system by modifying the governing variables, we examined the primary objectives and recommendations of the studies, in broad categories. Figure 8 present the bar graphs of each study's major objective (left) and recommendation (right). Left: the objectives were classified into 6 categories. Studies continue to introduce new syndromic surveillance systems (red bars), while there is also an increase in studies that attempts to improve existing methods (orange bars), and to test and extend the existing methods to new geographical regions (yellow bars). Right: recommendations are divided into three broad categories: advocative, optimistic or disapproving about the digital disease surveillance method in view. The proportions of each appear to remain rather steady over time. In earlier years, most studies aimed at 1) introducing new digital disease surveillance methods and 2) track public perception and reaction to influenza; in later years, more studies aimed at 3) improving and 4) comparing among existence methods, 5) extending these methods into new geographical regions, as well as ultimately 6) developing assimilated forecast models for influenza surveillance. The primary recommendations of the studies were roughly equally divided over time, with roughly half of the studies recommending the digital disease surveillance method in question over or as complement to traditional clinical surveillance methods, and the other half stating that it is promising yet awaits further research. This in part reflects the general endorsement of GFT by the research community as an acceptable real-time influenza surveillance tool, while working on improvements and searching for alternatives. It is worth noting that there were occasional but relatively high impact papers (e.g. Olson et al 2013) disapproving the usefulness of these methods, reflecting significant worries over their accuracies. According to our model, feedback loop 2 is at least partly driven by feedback loop 1. The timeline of the changes appears to coincide with such a hypothesis. The limitations being raised and improved appear to continue to support the acceptance of digital disease surveillance methods by the research community.

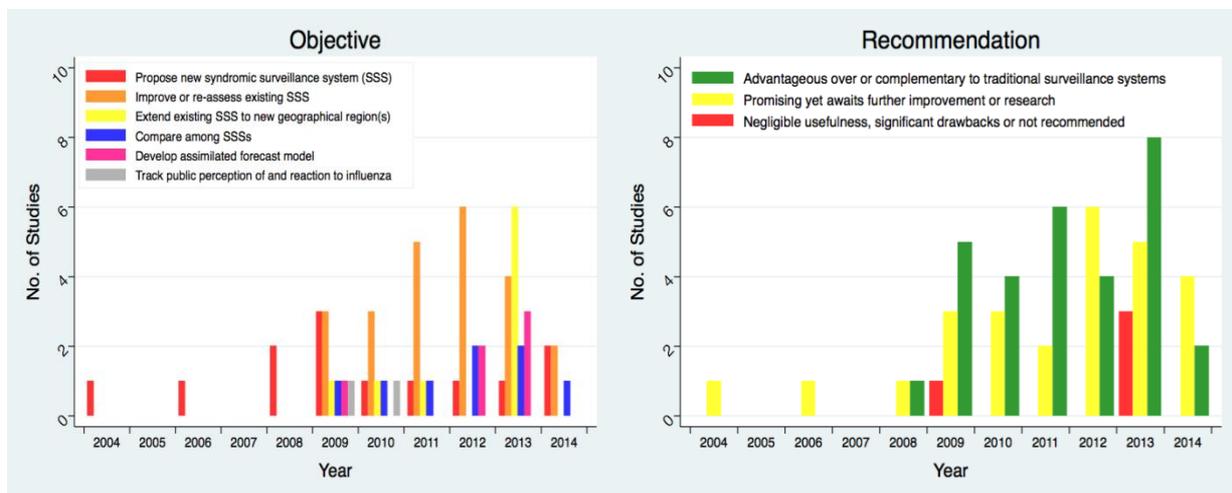

Figure 8. Bar graphs of each study's major objective (left) and recommendation (right)

**Conclusions**

We constructed a system dynamic model (SDM) for depicting and investigating into the co-evolutionary dynamics of digital disease surveillance algorithms in the past 10 years since the first research on the subject matter appeared in 2004. Quantifiable criteria were developed based on the PRISMA systematic review and meta-analysis guidelines, which allowed us to depict changes over time in study objectives, recommendations, approaches to accuracies, modeling, limitations raised and improved, etc.

We depicted system dynamic factors in terms of limitations raised and corrected ("Loop 1 learning"), as well as changes in the objectives and recommendations of the studies ("Loop 2 learning".) For loop 1: a significant portion of limitations raised were about confounding and noises, but network analysis suggests that over time the limitations became more specific to the kind of digital surveillance method being studied, together with a sharper increase in influences and collaborations during later years. In contrast, limitations improved at first tended to be specific to the kind of digital surveillance method being studied, and later more influences and collaborations appeared to be taking place, but with Twitter microblog studies in the centre, perhaps reflecting improvements in language processing and social modeling techniques. For loop 2: there were clear shifts over time in study objectives, from aiming to develop original digital disease surveillance methods and assessing public perception of and reactions to influenza, to the improvement, comparisons and assimilation of developed methods. Over time, there were almost equal and constant proportions of studies endorsing the digital surveillance method in view over or as complement to traditional surveillance methods, and of studies stating that the method in view is promising while awaiting further research. We suggest that this illustrates the general acceptance of digital disease surveillance methods by the research community, while the limitations being raised and improved continue to support such an acceptance.

As mentioned, we are working on results of the pooled correlations, errors and deviances in order to make more definite observations of the status of development of digital influenza surveillance. Further studies will also include the attempt to identify phase transitions and driving factors, as well as the dynamics of multi-disciplinary collaborations using behavioral correction models. Tentatively, our present novel methodology can be applied across digital surveillance of diseases other than influenza, and tentatively to other research areas, in characterizing their dynamic co‐evolutionary development.

**References**

Argyris, C., & Schon, D. (1978). Organizational learning: A theory of action approach. Reading, MA: Addision Wesley.
Brownstein, J. S., Freifeld, C. C., & Madoff, L. C. (2009). Digital disease detection—harnessing the Web for public health surveillance. New England Journal of Medicine, 360(21), 2153-2157.
Cho, S., Sohn, C. H., Jo, M. W., Shin, S. Y., Lee, J. H., Ryoo, S. M., ... & Seo, D. W. (2013). Correlation between national influenza surveillance data and google trends in South Korea. PloS one, 8(12), e81422.
Cook, S., Conrad, C., Fowlkes, A. L., & Mohebbi, M. H. (2011). Assessing Google flu trends Performance in the United States during the 2009 influenza virus A (H1N1) pandemic. PloS one, 6(8), e23610.
Copeland, P., Romano, R., Zhang, T., Hecht, G., Zigmond, D., & Stefansen, C. (2013). Google disease trends: an update. Nature, 457, 1012-1014.
de Lange, M. M., Meijer, A., Friesema, I. H., Donker, G. A., Koppeschaar, C. E., Hooiveld, M., ... & van der Hoek, W. (2013). Comparison of five influenza surveillance systems during the 2009 pandemic and their association with media attention. BMC public health, 13(1), 881.
Doornik, J. A. (2010). Improving the timeliness of data on influenza-like illnesses using Google Trends (Source: http://www.doornik.com/flu/Doornik%282009%29_Flu.pdf)
Dugas, A. F., Hsieh, Y. H., Levin, S. R., Pines, J. M., Mareiniss, D. P., Mohareb, A., ... & Rothman, R. E. (2012). Google Flu Trends: correlation with emergency department influenza rates and crowding metrics. Clinical infectious diseases, 54(4), 463-469.

Eysenbach, G. (2006). Infodemiology: tracking flu-related searches on the web for syndromic surveillance. In AMIA Annual Symposium Proceedings (Vol. 2006, p. 244). American Medical Informatics Association.

Ginsberg, J., Mohebbi, M. H., Patel, R. S., Brammer, L., Smolinski, M. S., & Brilliant, L. (2009). Detecting influenza epidemics using search engine query data. Nature, 457(7232), 1012-1014.

Johnson, H. A., Wagner, M. M., Hogan, W. R., Chapman, W., Olszewski, R. T., Dowling, J., & Barnas, G. (2004). Analysis of Web access logs for surveillance of influenza. Stud Health Technol Inform, 107(Pt 2), 1202-6.

Kang, M., Zhong, H., He, J., Rutherford, S., & Yang, F. (2013). Using Google trends for influenza surveillance in South China. PloS one, 8(1), e55205.

Lazer, D., R. Kennedy, G. King, and A. Vespignani. 2014. "The Parable of Google Flu: Traps in Big Data Analysis." Science 343 (6176) (March 14): 1203–1205.

Malik, M. T., Gumel, A., Thompson, L. H., Strome, T., & Mahmud, S. M. (2011). " Google flu trends" and emergency department triage data predicted the 2009 pandemic H1N1 waves in Manitoba. Canadian Journal of Public Health/Revue Canadienne de Sante'e Publique, 294-297.

Olson, D. R., Konty, K. J., Paladini, M., Viboud, C., & Simonsen, L. (2013). Reassessing Google Flu Trends data for detection of seasonal and pandemic influenza: a comparative epidemiological study at three geographic scales. PLoS computational biology, 9(10), e1003256.

Ortiz, J. R., Zhou, H., Shay, D. K., Neuzil, K. M., Fowlkes, A. L., & Goss, C. H. (2011). Monitoring influenza activity in the United States: a comparison of traditional surveillance systems with Google Flu Trends. PloS one, 6(4), e18687.

PRISMA (2009). http://www.prisma-statement.org/2.1.2%20-%20PRISMA%202009%20Checklist.pdf

Polgreen, P. M., Chen, Y., Pennock, D. M., Nelson, F. D., & Weinstein, R. A. (2008). Using internet searches for influenza surveillance. Clinical infectious diseases, 47(11), 1443-1448.

Thompson, L. H., Malik, M. T., Gumel, A., Strome, T., & Mahmud, S. M. (2014). Emergency department and 'Google flu trends' data as syndromic surveillance indicators for seasonal influenza. Epidemiology and infection, 142(11), 2397-2405.

Valdivia, A., Lopez-Alcalde, J., Vicente, M., Pichiule, M., Ruiz, M., & Ordobas, M. (2010). Monitoring influenza activity in Europe with Google Flu Trends: comparison with the findings of sentinel physician networks-results for 2009-10. Eurosurveillance, 15(29), 2-7.

Yuan, Q., Nsoesie, E. O., Lv, B., Peng, G., Chunara, R., & Brownstein, J. S. (2013). Monitoring influenza epidemics in china with search query from baidu. PloS one, 8(5), e64323.